\documentstyle[12pt,amssymb, epsfig]{article}
\input  epsf 
\def\rlx{\relax\leavevmode}
\def\inbar{\vrule height1.5ex width.4pt depth0pt}
\def\IZ{\rlx\hbox{\small \sf Z\kern-.4em Z}}
\def\IR{\rlx\hbox{\rm I\kern-.18em R}}
\def\ID{\rlx\hbox{\rm I\kern-.18em D}}
\def\IC{\rlx\hbox{\,$\inbar\kern-.3em{\rm C}$}}
\def\IN{\rlx\hbox{\rm I\kern-.18em N}}
\def\one{\hbox{{1}\kern-.25em\hbox{l}}}

\def\beq{\begin{equation}}
\def\eeq{\end{equation}}
\def\bea{\begin{eqnarray}}
\def\eea{\end{eqnarray}}
\def\ber{\begin{array}}
\def\eer{\end{array}}

\begin{document}

\begin{titlepage}

March 2000 \hfill{UTAS-PHYS-00-02}\\
\vskip 1.6in
\begin{center}
{\Large {\bf Integrity bases for local invariants of composite quantum 
systems:   }}\\[5pt]
{\Large  {\bf Corrigendum to J Phys {\bf A33} (2000) 1895-1914  }}\\[5pt]
\end{center}

\normalsize
\vskip .4in

\begin{center}
R I A Davis\footnote{
Dept of Physics, University of Queensland,
St Lucia Brisbane 4072}, \hspace{3pt} R Delbourgo, \hspace{3pt}
P D Jarvis
\par \vskip .1in \noindent
{\it School of Mathematics and Physics, University of Tasmania}\\
{\it GPO Box 252-21, Hobart Tas 7001, Australia }\\

\end{center}
\par \vskip .3in

\begin{center}
{\Large {\bf Abstract}}\\
\end{center}

\vspace{1cm}
Unitary group branchings appropriate to the calculation of local 
invariants of density matrices of composite quantum systems 
are formulated using the method of $S$-function plethysms.
From this, the generating function 
for the number of invariants at each degree in the density matrix 
can be computed.
For the case of two two-level systems, 
the generating function is
$F(q) = 1 +  q + 4q^{2} + 6 q^{3} + 16 q^{4} + 23 q^{5} + 52 q^{6}
+ 77 q^{7} + 150 q^{8} + 224 q^{9} + 396 q^{10} + 583 q^{11}+ O(q^{12})$.
Factorisation of 
such series leads in principle to the identification 
of an integrity basis of algebraically independent invariants.
This note replaces 
Appendix B of our paper\cite{us} J Phys {\bf A33} (2000) 1895-1914 
(\texttt{quant-ph/0001076}) which is incorrect.

\end{titlepage}

The measurement problem of detecting nonlocal differences between composite quantum
systems (for example, degrees of entanglement\cite{us}) 
is of great importance for applications to quantum computation and
communication. At root the question boils down to the identification 
of invariants with respect to unitary transformations which can be 
effected by local operations on each subsystem separately. For the 
case of two subsystems of dimensions $N_{1}$ and $N_{2}$, the  
$N_{1}N_{2} \! \times \!N_{1}N_{2}$ density matrix $\rho$ can be regarded in partition labelling\cite{Notation} as an element of the defining 
representation $\{ 1 \}$ 
 of $U({N_{1}}^{2}{N_{2}}^{2})$ branching to the 
reducible $\{ \bar{1} \} \{ 1 \} \! \times \!\{ \bar{1} \} \{ 1 \}$
representation of $U(N_{1}) \! \times \!U(N_{2})$ via 
$(\{ \bar{1} \} \! \times \!\{ 1 \}) \! \times \!(\{ \bar{1} \}\! \times \! \{ 1 \}) $
of $(U({N_{1}}) \! \times \!U({N_{1}})) \! \times \! (U({N_{2}} ) \! \times \!U({N_{2}} ) )$ within 
$\{ 1 \} \! \times \!\{ 1 \}$ of
$U({N_{1}}^{2}) \! \times \!U({N_{2}}^{2})$. Polynomial invariants of 
degree $n \ge 0$ are thus $U(N_{1}) \! \times \!U(N_{2})$ singlets of
the totally symmetric Kronecker power $\{1 \} \otimes \{ n \}
\equiv \{ n \} $ of $U({N_{1}}^{2}{N_{2}}^{2})$. According to the 
standard rules for plethysms\cite{Notation,Littlewood} the
branchings
  \begin{equation}
  U({N_{1}}^{2}{N_{2}}^{2}) \supset U({N_{1}}^{2}) \! \times \!U({N_{2}}^{2})
  \supset U(N_{1}) \! \times \!U(N_{2})	
  	\label{supsets}
  \end{equation}  
for this plethysm are
          \begin{eqnarray}
 \{ 1 \}  \otimes \{ n \} =  (\{ 1 \} \! \times \! \{ 1 \}) \otimes \{ n \}  
 &=&  \sum_{\sigma \; \vdash \; n} \; \{ \sigma \} \! \times \! \{ \sigma \circ n \} 
  \equiv  \sum_{\sigma \; \vdash \; n} \; \{ \sigma \} \! \times \! \{ \sigma  \} \nonumber \\
 & = & \sum_{
 \begin{array}{c} \kappa \; {\vdash }_{  N_{1} }\;  n, \\ 
 \lambda \; {\vdash }_{N_{2}} \; n \end{array} } \!\!
           \{ \bar{\kappa} \} \{ \kappa \circ \sigma \} \times
           \{ \bar{\lambda} \} \{ \lambda \circ \sigma \}. 
          \end{eqnarray}   
Here $\kappa$ and $\lambda$ must be $N_{1}$, $N_{2}$ part partitions 
of $n$ respectively in order that the corresponding representations of
$ U(N_{1}) $ and $ U(N_{2}) $ be nonvanishing. However, since the 
product of two representations in a unitary group will 
contain a singlet if and only if they are contragredient, the only singlets 
occurring are those for which both $\kappa \circ \sigma \ni \kappa $
and $\lambda \circ \sigma \ni \lambda$, or reciprocally
$ \sigma \in \lambda \circ \lambda$ and  $ \sigma \in \kappa \circ 
\kappa$. The number of singlets $F_{n}$ at degree $n$ is thus
   \begin{equation}
F_{n} = \sum_{
\begin{array}{c} \kappa \; {\vdash }_{  N_{1} }\;  n, \\ 
 \lambda \; {\vdash }_{N_{2}} \; n \end{array} }
   	n_{\kappa \lambda}
   	\label{}
   \end{equation}   
where $n_{\kappa \lambda}$ counts the number of $\sigma$ satisfying 
this condition\footnote{Also $\sigma$ should have at most 
$\mbox{min}({N_{1}}^{2},{N_{2}}^{2})$ parts.
In Appendix B of our paper\cite{us},
$n_{\kappa \lambda}$ was erroneously taken as 1 }. 
The computation thus reduces to the evaluation of 
inner products $\circ$ of $N_{1}$ and $N_{2}$ part partitions of $n$
(Kronecker products in the symmetric group $S_{n}$) and leads  
to the generating function
\begin{equation}
	F(q) = \sum_{n=0}^{\infty} F_{n}q^{n}.
		\label{}
\end{equation}
$F(q)$ is difficult to compute in closed form, but for specific 
cases can be evaluated to any desired degree\cite{Schur}. For example at 
degree 8 for the $2 \! \times \!2$ case, we find
\begin{eqnarray}
	\{ 6,2 \} \circ \{ 6,2 \} & =  & \{ 8 \} + \{ 7 1 \} + 2\{ 6 2 \} 
 +  \{ 6 1^{2} \} +  \{ 5 3 \} + 2\{ 5 2 1\} + \{ 5 1^{3} \} + \{ 4^{2} \} 
+ \{ 4 3 1\}  \nonumber \\
& & \mbox{} + \{ 4 2^{2} \};	\nonumber  \\
	\{ 5,3 \} \circ \{ 5,3 \} & = & \{ 8 \} + \{ 7 1 \} + 2\{ 6 2 \} 
  +  \{ 6 1^{2} \} +  \{ 5 3 \} + 2\{ 5 2 1\} + \{ 5 1^{3} \} + \{ 4^{2} \} 
  +  2\{ 4 3 1\}  \nonumber \\
  & & \mbox{} +  2\{ 4 2^{2} \}  + \{ 4 2 1^{2}\} + \{ 3^{2} 2\} + \{ 3^{2} 1^{2}\} 
    + \{ 3 2^{2}1\} ; 
	\label{}  
\end{eqnarray}
leading to a contribution (including multiplicity) of 
$n_{\{ 6,2 \},\{ 5,3 \} }=n_{\{ 5,3 \}, \{ 6,2 \}}= 18 $ to $F_{8}$. In this way we calculate
\begin{equation}
	F(q) = 1 +  q + 4q^{2} + 6 q^{3} + 16 q^{4} + 23 q^{5} + 52 q^{6}
+ 77 q^{7} + 150 q^{8} + 224 q^{9} + 396 q^{10} + 583 q^{11} + O(q^{12}).
	\label{Fseries}
\end{equation}
This generating function should be compared with the Molien 
series\cite{Grassl} defined via group integration,
\begin{equation}
	P(z) = \int_{g \in G} \frac{d\mu_{G}(g)}{\mbox{det}(\one - zg)}.
	\label{Molienseries}
\end{equation}
The equivalence between the two series can be readily established in the 
$S$-function formalism. Write $g$ as the element of $U(N_{1}^{2} N_{2}^{2})$
corresponding to the adjoint action of $U(N_{1}) \! \times \!U( N_{2})$ 
on $\rho$. Characters 
of group representations are generated by taking traces
$\langle g^{n} \rangle$ of 
powers of $g$, and hence are polynomials of the class parameters
$(x_{1}= \exp{i\phi_{1}}, x_{2}= \exp{i\phi_{2}}, \ldots )$. 
From the integrand of (\ref{Molienseries}) we have directly
\begin{eqnarray}
	{\left[{\mbox{det}(\one - zg)}\right]}^{-1} & = & 
	\prod_{i}\left. \frac{1}{(1- z x_{i})} \right.
	\label{} \nonumber \\
	& = &  \sum_{n=0}^{\infty} z^{n} S_{\{ n \}}(x)  
\end{eqnarray}
where a standard form of the Cauchy product identity 
has been used\cite{MacDonald} (the complete Schur functions $S_{\{ n \}}(x)$
correspond to one part partitions; for one argument $S_{\{ n \}}(z) = 
z^{n}$). The integrand at degree $n$ thus is 
indeed the reducible $U(N_{1}) \! \times \!U( N_{2})$ character
corresponding to the $n$'th symmetrised power ${\{ n \}}$ of the fundamental
representation of $U(N_{1}^{2} N_{2}^{2})$, and the 
invariant integration over $U(N_{1}) \! \times \!U( N_{2})$ serves to 
project the identity representation\footnote{
$F(q)$ in (\ref{Fseries}) agrees with $P(z)$ quoted by \cite{Grassl} 
up to degree 11}.

Makhlin\cite{Makhlin} has recently examined the $2 \! \times \!2$ case and 
proposed a concrete set of 18 local invariants for mixed state 
density matrices. Definitive confirmation of the completeness of such 
a set is in principle provided by a factorisation of $F(q)$ which
establishes an integrity basis presentation of the algebra of invariants 
in terms of a number of free generators
together with additional relations. Unfortunately (\ref{Fseries})
is not computed to sufficiently many terms to deduce an unequivocal factorisation, but 
for example the form
\begin{eqnarray*}
	G(x) & \equiv & \left. \frac{ (1+x^{4})(1+x^{5})(1+x^{6})^{4}(1+x^{7})^{2}
	(1+x^{8})^{2}(1+x^{9})^{2}}
	{(1-x)(1-x^{2})^{3}(1-x^{3})^{2}(1-x^{4})^{3}} \right.  \\
	 & = &  1+ x + 4 x^{2} + 6 x^{3}+ 16 x^{4} + 23 x^{5} + 52 x^{6} + \\
	&& \mbox{}+ 77 x^{7} + 150 x^{8} + 224 x^{9} + 398 x^{10} + 
	589 x^{11} + 982 x^{12} + O(x^{13})
		\label{Gseries} 
\end{eqnarray*}
may be noted. This has a denominator set signalling 9 free generators, 
including 1 at 
degree $1$ (the trace of $\rho$) and 3 at degree 2 (the traces of the 
squares of $\rho$ and of the reduced density matrices). The total count of 9 is to be expected from the dimensionality of 
the coset manifold $SU(4)/S(U(2)\! \times \!U(2))$, namely $8 = 15-(8-1)$ plus the overall 
singlet trace. 
To finite degree it is not possible uniquely to identify the 
denominator factors, and the set of free generators given by $G(x)$ differs from that 
implied by \cite{Makhlin}.
Of course, the saturation of terms in $G(x)$ in (\ref{Gseries}) compared with 
$F(q)$ in (\ref{Fseries}) beyond 
degree 9 also requires that some of the numerator factors (corresponding to invariant quantities 
whose squares or products are relations in the algebra)
should be combined together more economically. Nonetheless, the total 
count in $G(x)$ of 21 invariants (9 denominator plus 12 numerator 
quantities), and their degrees, is in agreement 
with \cite{Grassl}.

\subsubsection*{Acknowledgement}
The authors thank Prof Brian Wybourne for independently repeating our 
calculation and correcting an error.
Prof Ron King (private communication) has confirmed that the degree 12 coefficient in $F(q)$ 
is 964 in agreement with \cite{Grassl}. We further thank Dr Markus Grassl for 
correspondence about our results.

\end{document}